\documentclass[twocolumn,showpacs,aps,prl]{revtex4}

\usepackage{amssymb}
\usepackage{bm}
\usepackage{graphicx}
\usepackage[centerlast]{subfigure}

\newcommand{\rset}{{\mathbb{R}}} 

\newcommand{\tset}{{\mathbb{T}}} 
\begin{document}

\title{Topological Shocks in Burgers Turbulence}

\author{J.\  Bec$^1$}
\author{R.\  Iturriaga$^2$}
\author{K.\  Khanin$^{3\mbox{-}5}$}

\affiliation{$^1$ Lab.\ G.-D.\ Cassini, Observatoire de la C\^ote
  d'Azur, B.P.\  4229, 06304 Nice Cedex 4, France\\
  $^2$ Centro de Investigaci\'on en Matem\'aticas, A.P.\ 402,
  Guanajuato Gto., 36000 Mexico\\
  $^3$Isaac Newton Institute for Mathematical Sciences, 20
  Clarkson Road, Cambridge CB3 0EH, U.K.\\
  $^4$Department of Mathematics, Heriot-Watt University,
  Edinburgh EH14 4AS, U.K.\\
  $^5$ Landau Institute for Theoretical Physics, Kosygina Str.\ 2,
  Moscow 117332, Russia}

\draft{\textit{Phys.\ Rev.\ Lett.} \textbf{89}, 24501}

\begin{abstract}
  The dynamics of the multi-dimensional randomly forced Burgers
  equation is studied in the limit of vanishing viscosity. It is shown
  both theoretically and numerically that the shocks have a universal
  global structure which is determined by the topology of the
  configuration space. This structure is shown to be particularly
  rigid for the case of periodic boundary conditions.
\end{abstract}

\pacs{
47.27.Gs, 
02.40.Xx, 
05.45.-a  
}

\maketitle

\noindent The $d-$dimensional randomly forced Burgers equation
\begin{equation}
\partial_t {\bm u} + ({\bm u}\cdot \nabla){\bm u} = \nu \nabla^2 {\bm
  u} - \nabla F(\bm x,t),
\label{burg}
\end{equation}
appears in a number of physical problems, ranging from the dynamics of
interfaces and cosmology to hydrodynamics (see, e.g., Ref.\ 
\cite{houches} for a review). In the context of fluid dynamics, the
Burgers equation, frequently referred to as ``Burgers turbulence'', is
a simple model for analyzing the signature of singularities, mostly
shock discontinuities, in the statistics of the velocity field (see,
e.g., Refs.~\cite{cy95,p95,ekms,eve99,b01}). The (statistical)
steady-state theory for Burgers turbulence in the limit of vanishing
viscosity ($\nu \to 0$) was developed for $d=1$ in the spatially
periodic case \cite{ekms}.  The analysis of the Lagrangian dynamics
led to the distinguishing of a particular trajectory, the \emph{global
  minimizer}, corresponding to the unique fluid particle that is never
absorbed by a shock. The counterpart to the global minimizer is a
unique \emph{main shock}, with which all other shocks are merging
after a finite time and hence, in which all the matter gets
concentrated. Here the goal is to show that these objects, extended to
multi-dimensional situations, determine the global structure of the
stationary solution. This structure is strongly connected with the
topology of the configuration manifold defined by the boundary
conditions. We show that in any dimension, a unique global minimizer
exists, so that the shocks have either a local or a global topological
nature. The global shocks, unavoidably present in Burgers dynamics,
are called the \emph{topological shocks}; they have a nontrivial
structure for $d>1$. For simplicity, we mostly consider space-periodic
forcing potentials for which the configuration manifold is the
$d$-dimensional torus $\tset^d$ ($1$-periodic boundary conditions),
but most of our work can be extended to other types of  boundary
conditions and configuration spaces.

If the initial data is of gradient type, the velocity field preserves
this property at any later time, so that ${\bm u}({\bm x},t) = -\nabla
\psi({\bm x},t)$, where the velocity potential $\psi$ solves the
Hamilton--Jacobi equation
\begin{equation}
\partial_t \psi - \frac{1}{2}\left |\nabla \psi \right |^2 = \nu
\nabla^2 \psi + F({\bm x},t).
\label{hm}
\end{equation}
When the external potential $F$ is delta-correlated in both space and
time, this equation is known as the Kardar--Parisi--Zhang model for
interface dynamics~\cite{kpz86}. We focus here on smooth-in-space
forcing potentials with a correlation function given by
\begin{equation}
\left \langle F({\bm x}_1,t_1) F({\bm x}_2,t_2) \right \rangle = G({\bm
  x}_1-{\bm x}_2)\, \delta(t_1 - t_2),
\label{force}
\end{equation}
where $G$ is a smooth large-scale function. This type of large-scale forcing
was chosen analogous with that usually assumed in work on forced
Navier--Stokes turbulence. Note that the results discussed in this Letter can
be extended to other types of random forcing (e.g.\ with a finite correlation
time). Because of space periodicity, the average velocity ${\bm b} \equiv
\int_{\tset^d}{\bm u}({\bm x},t)d{\bm x}$ is an integral of motion. Its value
does not affect the structure of the topological shocks and, for simplicity,
we choose $\bm b=0$.

The initial-value problem associated to the Hamilton--Jacobi equation
(\ref{hm}), in the inviscid limit $\nu\to0$, has a variational
solution~\cite{variational}. Denoting $\psi_0$ the potential at the
initial time $t_0$, the velocity potential at times $t>t_0$ is given
by
\begin{equation} 
\psi(\bm x,t) = - \inf_{\bm\gamma(\cdot)} \left[ {\cal A} (\bm\gamma,
  t) - \psi_0(\bm \gamma(t_0)) \right],
\label{vissolution} 
\end{equation} 
where the infimum is taken over all differentiable curves $\bm\gamma:
[t_0,t] \to \tset^d$ such that $\bm\gamma(t)=\bm x$ and ${\cal A}$ is
the Lagrangian action
\begin{equation} 
{\cal A}({\bm\gamma},t)\equiv\int_{t_0}^{t} \left(\frac{1}{2} \left |
    \dot{\bm\gamma}(\tau) \right | ^2  -
  F({\bm\gamma}(\tau),\tau) \right)d\tau.
\label{action} 
\end{equation} 
A minimizing trajectory is called a minimizer on the interval
$[t_0,t]$ and is associated to a fluid particle reaching $\bm x$ at
time $t$. In the limit $t_0\to-\infty$, a stationary regime is
obtained, independent of $\psi_0$. The solution is then determined by
\emph{one-sided minimizers}, i.e.\ action-minimizing trajectories from
$-\infty$ to $t$. It is easily seen from Eq.\ (\ref{vissolution}) that
all the minimizers are solutions of the Euler--Lagrange equations
\begin{eqnarray}
\dot {\bm\gamma}(\tau) &=& \bm v(\tau),\label{eqX} \\
\dot {\bm v}(\tau) &=& -\nabla F(\bm \gamma(\tau),\tau). \label{eqV}
\end{eqnarray}
A \emph{global minimizer} (or a \emph{two-sided minimizer}) is defined
as a curve which minimizes the action for any time interval
$[t_1,t_2]$ and thus corresponds to the trajectory of a fluid particle
that is never absorbed by shocks.

We now state three main results whose rigorous proof is given in Ref.\ 
\cite{ik01} and that are essential for the introduction of topological
shocks. First, there exists a unique solution of the Hamilton--Jacobi
equation (\ref{hm}) in the limit $\nu\to0$ which is extendible to all
times. This solution is continuous and almost everywhere
differentiable in space.  It generates uniquely a stationary
distribution for the random Hamilton-Jacobi equation and its gradient
defines a unique statistically stationary solution of the inviscid
Burgers equation.  Second, for a given time and for every space
location where the potential is differentiable, there exists a unique
one-sided minimizer.  The locations where the one-sided minimizers are
not unique correspond to shock positions. Finally, there exists a
unique global minimizer.  This third statement is crucial for the
construction of topological shocks because it implies that for large
negative times $\tau\to-\infty$, all the one-sided minimizers are
asymptotic to the global minimizer \cite{remark2}.

\begin{figure}[ht]
  \centerline{\subfigure[\label{figspacetime}]
    {\includegraphics[width=6cm]{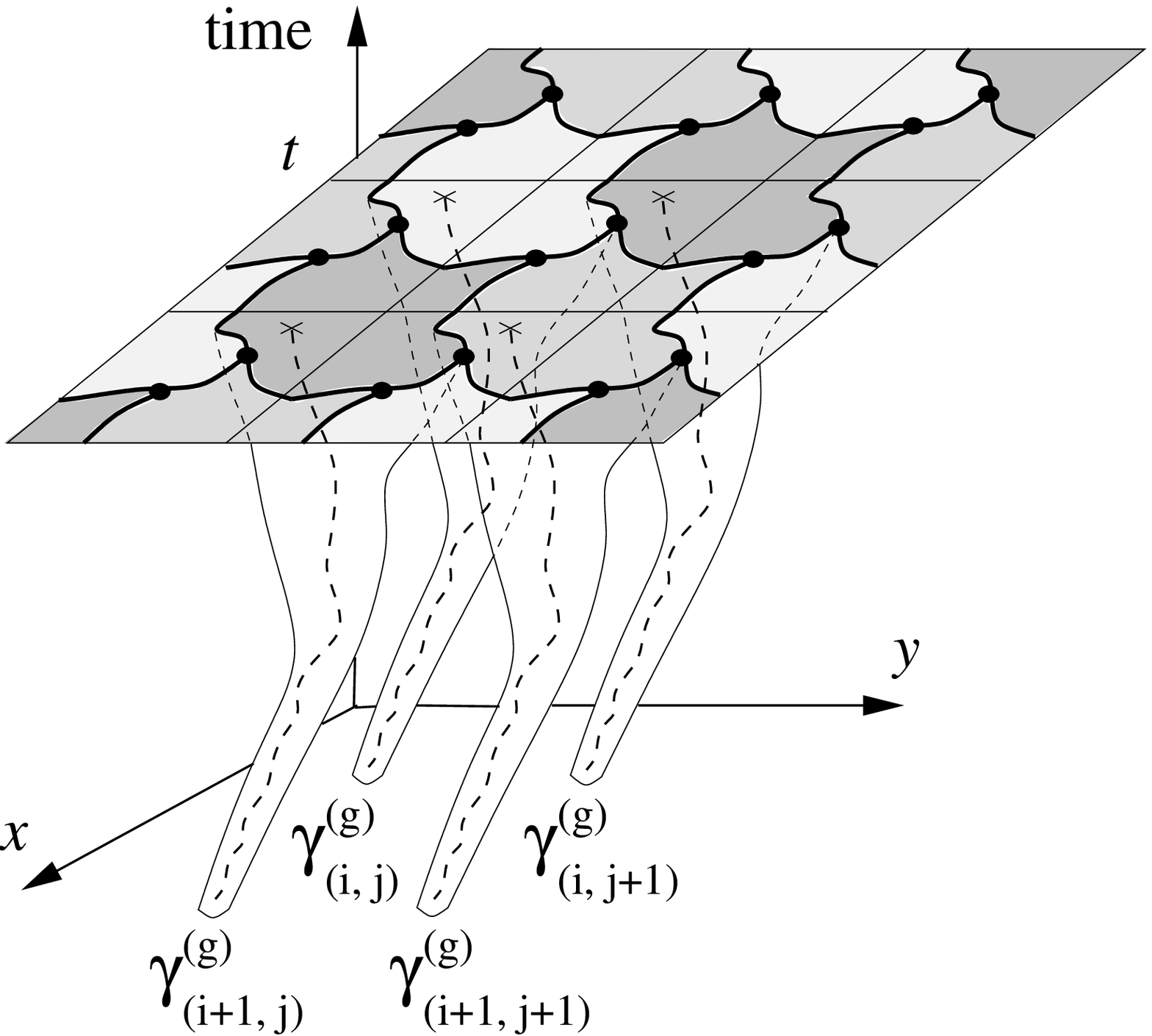}}
    \subfigure[\label{torus}]
    {\includegraphics[width=2cm]{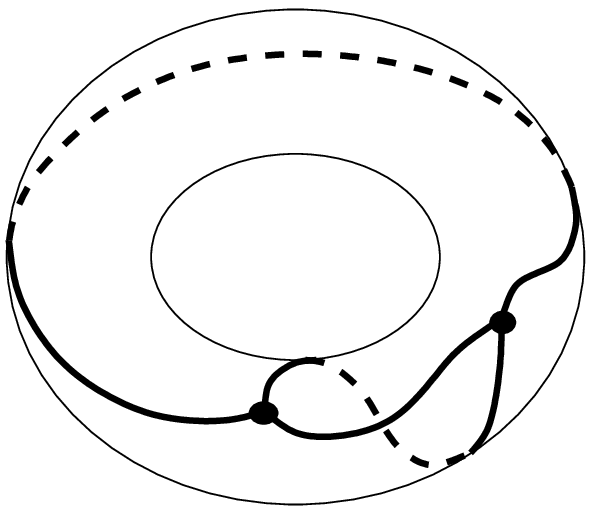}}}
  \caption{(a) Sketch in space-time of the unwrapped picture for
    $d=2$; on the horizontal plane which corresponds to a lattice of
    periodic box replicas at time $t$, the bold lines denote the
    topological shocks and triple points are represented by dots.  The
    different ``horns'' illustrate the backward-in-time convergence to
    four different global minimizers represented as dashed lines and
    the filled areas represent different tiles $O_{\bm k}$. (b)
    Position of the topological shock on the torus; the two triple
    points are represented as dots.}
\end{figure}

To introduce the notion of topological shock, we ``unwrap'', at a
given time $t$, the configuration space $\tset^2$ to the entire space
$\rset^d$ (see Fig.\ \ref{figspacetime}). Now, for a given realization
of the forcing, we obtain instead of a single global minimizer an
infinite number of them, each being the image of others by integer
shifts. They form a lattice parameterized by vectors $\bm k$ with
integer components and are denoted $\bm \gamma^{(\rm g)}_{\bm k}$. The
backward-in-time convergence to the global minimizer on $\tset^d$
implies that every one-sided minimizer emanating from some location in
$\rset^d$ is asymptotic to a particular global minimizer $\bm
\gamma^{(\rm g)}_{\bm k}$ on the lattice.  Hence, every location $\bm
x$ which has a unique one-sided minimizer is associated to an integer
vector $\bm k(\bm x)$, defining a tiling of the space at time $t$. The
tiles $O_{\bm k}$ are the sets of points whose associated one-sided
minimizer is asymptotic to the $\bm k$-th global minimizer.  The
boundaries of the $O_{\bm k}$'s correspond to the positions of
particular shocks that are called the topological shocks. They are the
locations for which at least two one-sided minimizers approach
different global minimizers on the lattice.  Indeed, a point where two
tiles $O_{\bm k_1}$ and $O_{\bm k_2}$ meet, has at least two one-sided
minimizers, one of which is asymptotic to $\bm \gamma^{(\rm g)}_{\bm
  k_1}$ and another to $\bm \gamma^{(\rm g)}_{\bm k_2}$. Of course,
there are also points on the boundaries where three or more tiles meet
and thus where more than two one-sided minimizers are asymptotic to
different global minimizers.  For $d=2$ such locations are isolated
points corresponding to the intersections of three or more topological
shock lines, while for $d=3$, they form edges and vertices where shock
surfaces meet. Note that, generically, there exist other points inside
$O_{\bm k}$ with several minimizers.  They correspond to shocks of
``local'' nature because at these locations, all the one-sided
minimizers are asymptotic to the same global minimizer $\bm
\gamma^{(\rm g)}_{\bm k}$ and hence, to each other. In terms of mass
dynamics, the topological shocks play a role dual to that of the
global minimizer. Indeed, all the fluid particles are converging
backward-in-time to the global minimizer and are absorbed
forward-in-time by the topological shocks. Assuming that the Burgers
equation (\ref{burg}) is accompanied by a continuity equation for the
mass density, this implies that all the mass concentrate at large
times in the topological shocks.

\begin{figure}[ht]
  \centerline{\includegraphics[width=7.5cm]{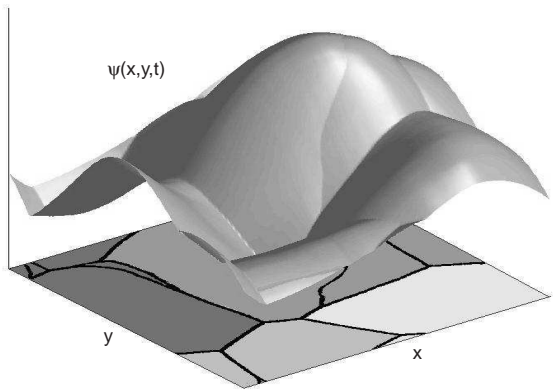}}
  \caption{Snapshot of the velocity potential
    $\psi(x,y,t)$ for $d=2$ in the statistical steady state, obtained
    numerically with $256^2$ grid points. Shock lines, corresponding
    to locations where $\psi$ is not differentiable, are represented
    as black lines on the bottom of the picture; the four gray areas
    are different tiles separated by the topological shocks; the other
    lines are local shocks.}
  \label{potential}
\end{figure}

We now describe the global structure of the topological shocks.
Parameter counting suggests that there are generically
$(d-1)$-dimensional surfaces of points with two one-sided minimizers
which contain $(d-2)$-dimensional sub-manifolds with three one-sided
minimizers and so on. This ends up with single points (zero-dimension)
from which emanate $(d+1)$ one-sided minimizers. As one expects to see
only generic behavior in a random situation, the probability to have
points associated to more than $(d+1)$ one-sided minimizers is zero.
It follows that there are no points where $(d+2)$ tiles $O_{\bm k}$
meet, an important restriction on the structure of the tiling. Thus
for $d=2$, the tiling is constituted of curvilinear hexagons. Indeed,
suppose each tile $O_{\bm k}$ is a curvilinear polygon with $s$
vertices corresponding to triple points.  For a large piece of the
tiling which consists of $N$ tiles, the total number of vertices is
$n_v \sim sN/3$ and the total number of edges is $n_e \sim sN/2$.  The
Euler formula implies that $1 = n_v - n_e + N \sim (6-s)N/6$; so we
have $s=6$, corresponding to an hexagonal tiling. As shown in
Fig.~\ref{torus}, this structure corresponds on the periodicity torus
$\tset^2$, to two triple points connected by three shock lines which
are the curvilinear edges of the hexagon $O_{\bm 0}$. The connection
between the steady state velocity potential and the topological shocks
is shown on Fig.~\ref{potential} which was obtained numerically.

Although the topological shocks always form hexagonal patterns when
$d=2$, the corresponding tilings can be of different types; in the
course of time, the merger of two triple points is the generic
mechanism for changing the type of the tiling. This so-called flipping
bifurcation~\cite{AppendixBookGurbatov} has the property of
redistributing matter among nodes, so that the mass does not
concentrate in a particular triple point. In higher dimensions, the
structure of topological shocks can be more complex; for instance, it
is not possible to determine uniquely the shape of polyhedra forming
the tiling. Nevertheless, the minimal polyhedra defining such tilings
for $d=3$ can be shown to have 24 vertices \cite{anisov,matveev}.

All the above results concerning the global structure of solutions
require a statistical steady state, achieved asymptotically at large
times. The convergence to this regime is actually exponential so that,
generally, the global picture of the flow is reached after just a few
turnover times.  The nature of the convergence is related to the local
properties of the global minimizer and more particularly to its
hyperbolicity. For $d=1$, the global minimizer has been shown to be a
hyperbolic trajectory of the dynamical system defined by the
Euler--Lagrange equations (\ref{eqX}-\ref{eqV}) \cite{ekms}. In
multi-dimensional situations the hyperbolicity of the global minimizer
is an open problem. Since the Lagrangian flow defined by
(\ref{eqX}-\ref{eqV}) is Hamiltonian, one can define $d$ pairs of
non-random Lyapunov exponents with opposite signs.  Hyperbolicity
means that none of these exponents vanishes.  This question can be
addressed in terms of the backward-in-time convergence of the
one-sided minimizers to the global one or, in terms of forward-in-time
dynamics, by looking at how fast Lagrangian fluid particles are
absorbed into shocks.  For this, we consider the set $\Omega(T)$ of
locations $\bm x$ such that the fluid particle at $\bm x$ at time
$t=0$ survives, i.e.\ is not absorbed by any shock, until the time
$t=T$. The long-time shrinking of $\Omega$ as a function of time is
asymptotically governed by the Lyapunov exponents. To ensure the
absence of vanishing Lyapunov exponents, it is sufficient to show that
the diameter of $\Omega(T)$ decays exponentially as $T \to \infty$.
Below we demonstrate numerically that this is indeed the case for
$d=2$.
\begin{figure}[ht]
  \centerline{\includegraphics[width=8.5cm]{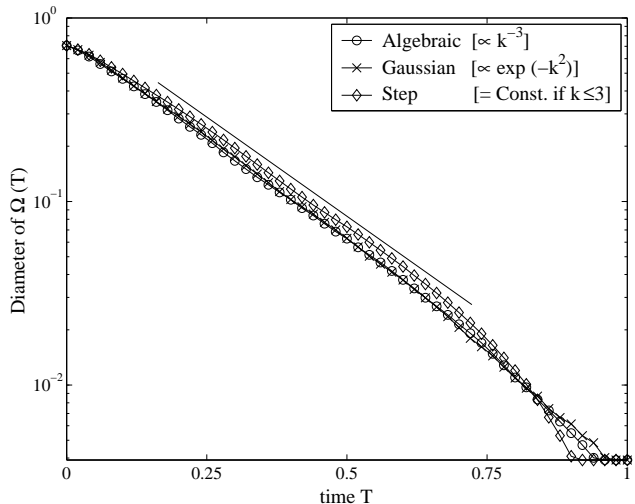}}
  \caption{Time evolution of the diameter of the Lagrangian set
    $\Omega(T)$ (points corresponding to the regular region) for three
    different types of forcing spectra normalized to give the same
    Lyapunov exponents; average over 100 realizations and with $256^2$
    grid points.}
  \label{hyperbolicity}
\end{figure}
For this we assume that the forcing is a sum of independent random
impulses concentrated at discrete times \cite{bfk00}, a case to which
the present theory remains applicable. Between ``kicks'' the velocity
field decays according to the unforced Burgers equation. This allows
us to use the fast Legendre transform method \cite{nv94}, based on
discrete approximations of the one-sided minimizers over a grid, which
gives directly the solution in the inviscid limit and is particularly
useful due to its strong connection with the Lagrangian picture of the
flow. We can then track numerically the set $\Omega(T)$ of regular
Lagrangian locations. As shown in Fig.~\ref{hyperbolicity} for three
different types of forcings, the diameter of this set decays
exponentially in time, providing strong evidence for the
hyperbolicity of the global minimizer when $d=2$.

Since all the one-sided minimizers converge backward-in-time to the
global minimizer, hyperbolicity implies that, in the statistical
steady state, the graph of the solution in the phase space $({\bm x},
{\bm u})$ is made of pieces of the smooth unstable manifold associated
to the global minimizer with discontinuities along the shocks lines or
surfaces. In other words, shocks appear as jumps between two different
folds of the unstable manifold. The smoothness of the unstable
manifold is key; for instance, it implies that when $d=2$, the
topological shock lines are smooth curves. The above geometrical
construction of the solution has much in common with that appearing in
the unforced problem. Indeed, when $F=0$, the solution can be obtained
by considering in the $({\bm x}, {\bm u})$ space, the Lagrangian
manifold defined by the position and the velocity of the fluid
particles at a given time. This analogy gives good ground to
conjecturing that several universal properties associated to the
unforced problem still hold in the forced case, as indeed happens in
one dimension \cite{ekms,b01}. Hyperbolicity implies that there exists
a strong parallel between the forced and the unforced situations.
Despite the fact that the mass dynamics is completely different in the
two cases (mass is absorbed by shocks linearly in time in the decaying
case and exponentially fast in the forced case), many features of the
velocity field are universal.  For instance, it was shown in
Ref.~\cite{fbv01} that, for the unforced case and $d>1$, large but
finite mass densities are localized near time-persistent boundaries of
shocks (``kurtoparabolic'' singularities) contributing, in any
dimension, power-law tails with the exponent $-7/2$ in the probability
density function (PDF) of both velocity gradients and mass densities.
When a force is applied, the geometry of the solutions is very similar
to that appearing in unforced situations. This leads again to a
universal $-7/2$ power-law behavior of the PDF of velocity gradients
and mass density, irrespective of $d$. The issue of similarities
between forced and unforced Navier--Stokes turbulence and the search
for universal statistical properties of the velocity field is, of
course, still an open problem.

Note, finally, an important physical problem which will be addressed in a
future work, namely the understanding of the behavior of minimizers and the
effects related to global shocks in the case of \emph{spatially extended
  non-periodic systems}. This amounts to investigating intermediate time
asymptotics when the size of the system is much larger than the forcing scale.
Preliminary numerical results indicate the appearance of a shock structure
resembling the structure of topological shocks.

\noindent We are grateful to S.\ Anisov, I.\ Bogaevski, U.\ Frisch,
T.~Matsumoto, S.\ Novikov, Ya.\ Sinai, S.\ Tarasov for illuminating
discussions and useful remarks. R.I.\ was supported by
CONACYT-M{\'{e}}xico grant 36496-E. Simulations were performed in the
framework of the SIVAM project of the Observatoire de la C\^ote
d'Azur, funded by CNRS and MENRT.


\begin{thebibliography}{99}
\bibitem{remark2} Although there are no doubts that backward-in-time
  convergence of one-sided minimizers to the global one always holds,
  we were able to prove it rigorously only for a non-random sequence
  of times $\tau_n \to -\infty$. Notice, however, that convergence for
  a sequence is enough for the construction of the topological shocks.
\end{thebibliography}
\end{document}